# Evaluating Learning of Motion Graphs with a LiDAR-Based Smartphone Application

**Abstract:**
Data modeling and graphing skill sets are foundational to science learning and careers, yet students regularly struggle to master these basic competencies. Further, although educational researchers have uncovered numerous approaches to support sense-making with mathematical models of motion, teachers sometimes struggle to enact them due to a variety of reasons, including limited time and materials for lab-based teaching opportunities and a lack of awareness of student learning difficulties. In this paper, we introduce a free smartphone application that uses LiDAR data to support motion-based physics learning with an emphasis on graphing and mathematical modeling. We tested the embodied technology, called *LiDAR Motion*, with 106 students in a non-major, undergraduate physics classroom at a mid-sized, private university on the U.S. East Coast. In identical learning assessments issued both before and after the study, students working with *LiDAR Motion* improved their scores by a more significant margin than those using standard issue sonic rangers. Further, per a voluntary survey, students who used both technologies expressed a preference for *LiDAR Motion*. This mobile application holds potential for improving student learning in the classroom, at home, and in alternative learning environments.

**Authors**

| | | |
|---|---|---|
| Daniel J. O'Brien | Rebecca E. Vieyra | Chrystian Vieyra Cortés |
| Georgetown University | Vieyra Software | Vieyra Software |
| Washington, DC | Washington, DC | Washington, DC |

Mina C. Johnson-Glenberg
Arizona State University
Tempe, AZ

Colleen Megowan-Romanowicz
American Modeling Teachers Association
Sacramento, CA

## I. Introduction

Although time-based motion graphs are fundamental to teaching and learning kinematics, students' intuitions about how to interpret them are often misaligned with what graphs actually represent (Esach, 2014). That students struggle to read graphs has been well-documented since the earlier days of physics education research (Clement, 1985; Goldberg & Anderson, 1989l; McDermott et al., 1987; Trowbridge & McDermott, 1980). The resulting taxonomies of student difficulties on this topic became the basis for heavily-used concept inventories, such as the Mechanics Baseline Test (Hestenes & Wells, 1992) or the Test of Understanding Graphs in Kinematics (Beichner, 1994).

Educators have proposed a number of pedagogical and technological supports to help students make sense of motion graphs. Early on, educators conceived of microcomputer laboratories with commercially-designed, specialized ultrasonic rangers attached to a computer visualization output (Thornton, 1987; Thornton & Sokoloff, 1990; Reddish et al., 1997). Seeking other technologies that are more accessible, versatile, or cheaper, recent proposals have incorporated ultrasonic rangers on Arduino boards (Ladeira et al., 2022), robots (Sena dos Santos & Mendes dos Santos, 2020; Balaton et al., 2020), and Kinect for *Xbox One* (Anderson & Wall, 2016). Despite the potential of tools such as mobile handheld devices for collecting and visualizing physics-related data for many topics (Vieyra et al., 2015), the low-cost mobile accelerometers embedded within them are generally not adequate for capturing accurate position data—at least in introductory physics contexts—due to a need for regular recalibration and data cleaning (Pang & Liu, 2001).

The addition of light detection and ranging (LiDAR) technology to some models of iPads in March 2020 and iPhones in October 2020 provided a radical new opportunity to measure motion with high-precision position sensors. Through our creation of a new app, iPad and iPhone users now have the capability to visualize graphs of their own motion in real time, providing new opportunities to understand kinematics using technology that many learners already have in their pockets. Research supports that when the immediacy of motion visualization is paired with a learner's own motion there are significantly positive learning outcomes (Duijzer et al., 2019). This paper presents the outcome of an introductory-level university physics study using the *LiDAR Motion* app (www.lidar-motion.net) for LiDAR-based iPads and iPhones. It demonstrates the utility of this new technology, especially in light of many teachers' need to flexibly respond to circumstances such as distance learning resulting from the COVID pandemic (O'Brien, 2021).

**II. Technology for Quantifying Linear Motion**

This study tested the usefulness of *LiDAR Motion*, a free mode within the *Physics Toolbox Sensor Suite* application available for iPad and iPhone. It was developed to plot a user's movement as they move directly perpendicular to a flat wall or object (Figure 1). The LiDAR hardware in the mobile device emits an array of infrared laser pulses and then senses the reflected pulses, using the light's time of flight to calculate the distance from the phone to the object it is pointing toward. The resulting information is used by the app to display real-time data on position-time and/or velocity-time graphs on the screen (Figure 2), allowing users to understand how the graphs coordinate with their movements. Additionally, the app includes a game challenge feature in which users must attempt to match their motion to pre-drawn graphs (Figure 3). Each challenge was developed to address common student difficulties, and users are only successful once their motion data is validated against expected values within an acceptable $R^2$ threshold. When the user has moved in a manner that stays within a prescribed error zone around the displayed graph, the app displays a congratulatory "Great job!" message with confetti.

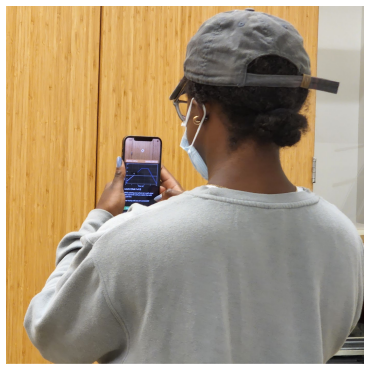

**Figure 1.** A student uses *LiDAR Motion* as she moves perpendicular to the wall, toward and away from it.

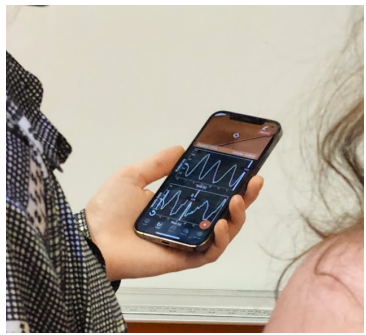

**Figure 2.** Students reflect on a sinusoidal movement pattern collected with *LiDAR Motion*.

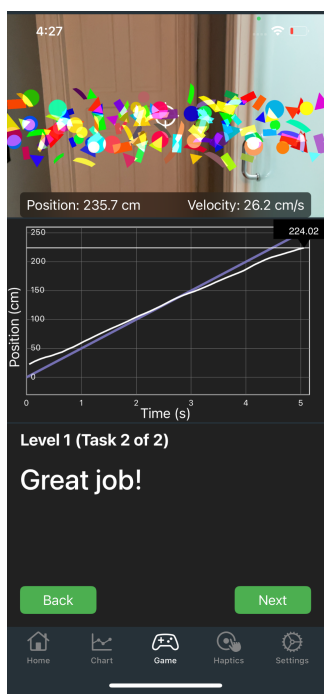

**Figure 3.** The app interface showing a pre-drawn motion graph (purple), actual user data produced by human motion (white), and confetti to congratulate the user for achieving a good fit with the pre-drawn graph

## III. Research Questions

Considering the increasing access that teachers and learners have to mobile devices, the purpose of this study was to determine whether physics students learn about motion differently when using the newly developed *LiDAR Motion* application as compared with specialized commercial science equipment: an ultrasonic ranger, track, cart, and computer graphical output. Understanding the affordances of these technologies within the same inquiry-based pedagogical approach will shine a light on how teachers might prioritize technology selection for teaching motion graphs.

## IV. Method

The technology was tested in a non-major, introductory physics course at a mid-sized private university on the U.S. East Coast. Eight laboratory sections were randomly assigned to either the intervention or the control: four groups were assigned to use *LiDAR Motion* and four groups were assigned to use *Vernier* ultrasonic rangers paired with *LoggerPro*. Regardless of technology, students were placed into pairs and asked to take turns as they carried out laboratory tasks.

The laboratory sections were led by the course's regular instructor and supported by at least two undergraduate or graduate teaching assistants, all of whom were independent of this project's research team. Additionally, one of the researchers met with the course instructor and the teaching assistants a week prior to the laboratory sessions to introduce them to the phone application and to the *Vernier* equipment and software. At least one researcher was present during each of the laboratory sessions to observe student interactions.

Students' understanding of kinematic concepts was evaluated quantitatively using a written assessment. The assessment was composed of fifteen questions about position and velocity graphs sampled from the Test of Understanding Graphs in Kinematics v4.0 ("TUG-K," Zavala et al., 2017) and Diagnoser (Minstrell, 2003). The test was administered to students one week prior to the laboratory session (pre-test) and immediately following the laboratory session (post-test). The pre- and post-test results were compared against each other and between the intervention and control groups.

Students in each section were provided with nearly identical protocol sheets to guide them through the laboratory session (Figure 4). The only differences pertained to specific instructions about the technology they were using (see: protocol sheets in Supplemental Documents, Annex A). The protocol sheets began with a free play session during which the students could explore the technology to which they were assigned. Subsequently, students were guided into a set of tasks to challenge their understanding of kinematics and motion mapping. This feature is built into *LiDAR Motion* as a graph-matching game in which the user walks perpendicular to a wall to match the shape of a position-time curve to their movement. Students using the ultrasonic detector were instructed to use their bodies to walk in front of the detectors (Figure 5) to create position-time motion graphs drawn on the protocol sheet—which mirrored those in the *LiDAR Motion* graph match game. All sections completed the protocol within 45-60 minutes, regardless of the technology they had been assigned.

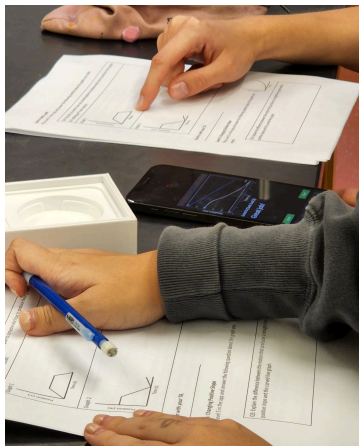 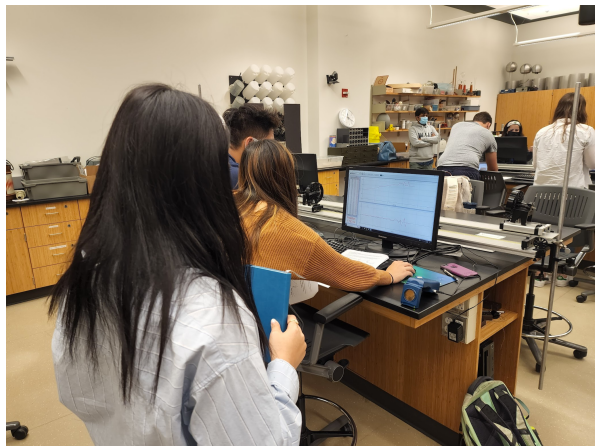

**Figure 4.** Students reflect on the graphs they have just matched to determine how their motion corresponds to aspects of the graphs.

**Figure 5.** Two students observe *LoggerPro* while a third student walks out a motion graph in front of an ultrasonic ranger.

Lastly, one separate laboratory section employed both forms of technology, with half of the students starting with *LiDAR Motion*, the other half of the students starting with ultrasonic rangers, and both groups switching halfway through. Students in this section were also asked to voluntarily complete an interest form after the laboratory session. This form gauged students' impressions of the level of enjoyment of the activities, the level of usefulness of the activities to learning the concepts, the level of engagement of the activities, and which technology they preferred using: *LiDAR* or ultrasonic. The form also had a free-response prompt for students to comment on the activities.

**V. Results**

The written assessment test was used to gauge the effectiveness of *LiDAR Motion* relative to a standard classroom technology for the eight sections that completed the full protocol with a single technology. The maximum score on the test was 15, and pre-test scores in all sections were relatively high (Figure 6a). The two groups (*LiDAR* and ultrasonic) were matched at pretest, $t(104) = 1.33$, $p = 0.19$. By post-test, the *LiDAR Motion* group increased by a statistically significant 1.32 points, $t(56) = 4.12$, $p < 0.001$, and the ultrasonic group improved by a statistically significant 0.53 points, $t(48) = 2.65$, $p < 0.05$. Per an ANOVA comparing these gains (post - pretest), the increase in the LiDAR group's mean score after intervention was greater than that of the sonic group's at a statistically significant level, $F(1, 104) = 4.02$, $p < .05$.

The number of students who answered correctly on each assessment question was also compared. Students in the *LiDAR Motion* sections improved more on twelve of the fifteen questions (Figure 6b). Raw data illustrating the distribution of scores are available (see: Supplemental Documents, Annex B).

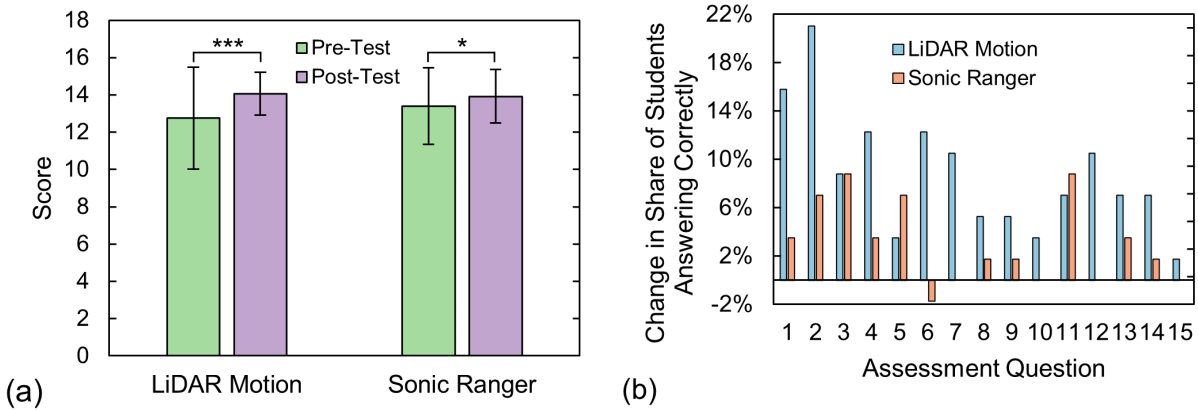

**Figure 6.** (a) Comparison of students' pre- and post-intervention learning assessment scores (out of 15 points; ***$p<0.001$, *$p<0.05$*). (b) Questions-by-question change in students' scoring on the assessment.

Students in the ninth laboratory section who completed tasks with both technologies were asked which technology they preferred. Of the fourteen individuals who responded, two students preferred the ultrasonic ranger, eleven students preferred *LiDAR Motion*, and one had no preference (Figure 7a). Those who preferred *LiDAR Motion* rated the activities' level of enjoyment and engagement more favorably on a scale from 1 to 5 (Figure 7b). Given the small number of students who preferred the ultrasonic ranger or had no preference, inferential statistics were not calculated.

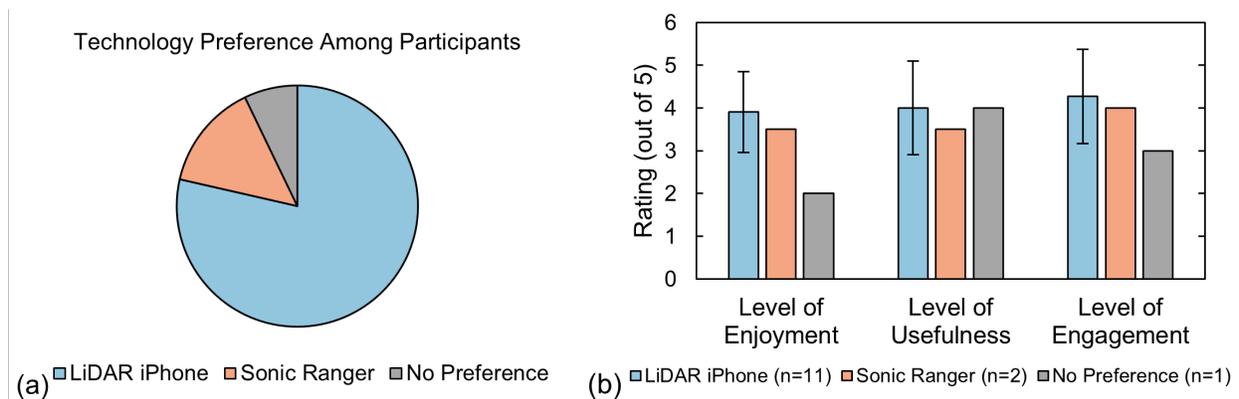

**Figure 7.** Results from follow-up survey. (a) Chart of students' technology preference. (b) Students' evaluation of the laboratory activities categorized by which technology they preferred.

Qualitatively, researchers observed the nature of student interactions with the technology during the laboratory sessions. Students were on task the majority of the time in both the treatment and control laboratory sessions, and all pairs were highly collaborative. However, those using the app expressed more frequent and more intense emotions, including fist bumps, high fives, and exclamations upon achieving a graph match. One student shared with her lab partner, "It's actually kind of fun. I wish I had this in high school." Many students also expressed a sense of anxiousness about doing the graph match tasks: "My heart is beating so fast," and "It makes me so nervous. It's a good nervous…pressure to get it right." Some students expressed frustration about not being able to match the graph: "I know what I'm supposed to be doing, but my mind and my feet aren't coordinating."

Although tasks in the *LiDAR* and ultrasonic ranger protocols were designed to mirror each other—requiring students to walk linearly to replicate a particular graph shape—students sometimes deviated from the written protocol. For example, multiple groups using ultrasonic rangers instead used their arms to move a book along a track in front of the detector to make graphs, detracting from the intended locomotive embodiment of the activity (Figure 8). Moving the arm is considered a lower level of embodiment compared to walking which activates more sensori-motor neurons, among other systems (Johnson-Glenberg, 2018). The real-time response and handheld nature of the iPhone app kept students from deviating in a similar way from the protocol while using *LiDAR Motion*, though at least one student was seen moving the phone towards and away from the wall by extending and retracting their arms, rather than by walking.

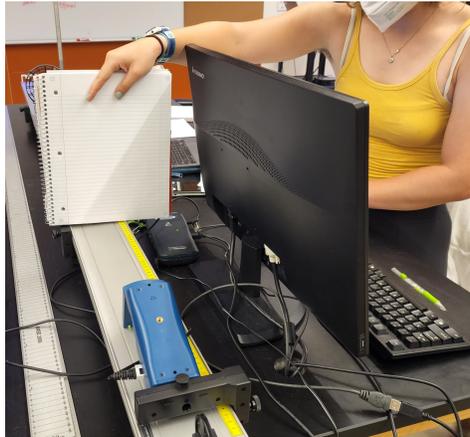

**Figure 8.** A student moves a notebook in front of their ultrasonic ranger rather than "walking out" the graph.

## VI. Discussion and Conclusion

This research study illustrates that *LiDAR Motion* results in more significant achievement gains for undergraduate non-major students when compared to those who used commercial ultrasonic rangers and associated software. The question-by-question results offered a more granular perspective on learning gains. For example, in contrast with those using ultrasonic rangers, students using *LiDAR Motion* improved most significantly on assessment questions 1, 2, 6, and 7; these four questions all relate to single-segment analyses of constant velocity. Although one can attempt to match a graph on *LoggerPro*, the user is not given real-time feedback on accuracy. The feedback on graph match goodness of fit in *LiDAR Motion* may contribute more to students' comprehension of motion, especially for simpler graphs. The app forces students to re-evaluate and re-attempt their movements until it validates their graph match with positive feedback. Matching graphs using the whole body and doing that task multiple times until "correct" may have led to the increase in gain scores for the LiDAR group.

Contrary to the belief held by some that smartphones may pose a distraction in learning environments, our own study and related studies by Kaps et al. (2021) and Mazeella and Testa (2016) showed that mobile devices can be highly effective teaching tools. Further, we observed a strong preference for the use of *LiDAR Motion*, which echoes findings by Hochberg et al. (2018) and Ozkan (2016) that students' enjoyment, interest, and curiosity in science increase when using mobile devices.

One notable aspect of this study is that students displayed highly collaborative engagement, with or without the use of the smartphone. Prior to the study, the researchers had some concerns that the use of a personal mobile device, with its small screen, might threaten students' cooperative actions. However, these concerns were unfounded, as students in the intervention and control groups could be seen discussing their planned motions and coaching each other. This finding stands in contrast to Anderson and Wall (2016), in which the use of *Xbox One* decreased student collaboration—although this was potentially a result of the fact that they used a single piece of technology with the entire class, while the present study had one set of technology for each pair of students.

Our findings are limited by a number of factors that can be mostly attributed to the advanced nature of student understanding. Although the research participants were introductory non-major physics students, the highly competitive nature of the university suggests that many of these students had prior physics classes in high school and had already developed good graph reading skills. Some students also mentioned that they had previously used sonic rangers. The effect was that students had substantial prior knowledge, and pre-test scores did not reflect an expected normal distribution (see: Annex B). Future studies should consider working with a more general audience or students much earlier in their learning trajectory, perhaps including secondary students who are just learning algebra (around ages 12-14). Alternatively, more difficult questions could allow a deeper understanding of upper-level students' learning gains with the technology.

The present study is particularly applicable to schools that have policies that include iPads or permit the use of personal devices, but may not have the budget for specialized science equipment. It is also applicable in cases where students need to learn from home. Our study places tools such as iPads and iPhones with *LiDAR Motion* among the set of viable options from which educators can select to effectively teach about motion graphs.

**Acknowledgments**
This research was supported by the National Science Foundation Grant #2114586 and approved by Georgetown University Institutional Review Board (Study00005454). Special thanks to Christopher Cothran and the physics teaching assistants of Georgetown University. Special appreciation to Thérèse Vieyra, who assisted in the development and testing of this activity.